\documentclass[twocolumn,showpacs,preprintnumbers,amsmath,amssymb,floatfix]{revtex4-1}
\usepackage[perpage,symbol]{footmisc}
\usepackage{color}

\setfnsymbol{bringhurst}

\normalsize 
\usepackage{graphicx}
\usepackage{epsfig} 
\usepackage{dcolumn}
\usepackage{bm}
\usepackage{epstopdf}

\begin{document}

\preprint{APS/123-QED}

\title{Gain measurement scheme for precise determination of atomic parity violation through two-pathway coherent control}

\author{J. Choi$^{1,3}$, R. T. Sutherland\footnote{Current address: Lawrence Livermore National Lab, 
7000 East Avenue, Livermore, CA, 94550}
$^{2}$, George Toh$^{1,3}$, A. Damitz$^{2}$ and D. S. Elliott$^{1,2,3}$ }
\affiliation{%
  $^1$School of Electrical and Computer Engineering, $^2$Department of Physics and Astronomy, $^3$Purdue Quantum Center \\ Purdue University, West Lafayette, IN  47907 
}

\date{\today}

\begin{abstract}
Precision measurements of parity non-conserving (PNC) interactions in atoms, molecules and ions can lead to the discovery of new physics beyond the standard model and understanding of weak-force induced interactions in the nucleus. 
In this paper, we propose and analyze a novel atomic parity violation measurement scheme for a forbidden transition 
where we combine a two-pathway coherent control mechanism with probe gain techniques. 
We detail a feasible experimental geometry for $6S_{1/2}\rightarrow 7S_{1/2}$ transitions in a cesium vapor cell, and consider the statistical noise of such a measurement under reasonable laboratory conditions.
We estimate the signal-to-noise ratio to be approaching $\sim2.3/\sqrt{Hz}$.
This scheme, with low expected systematic errors, would allow for precise measurements in cesium and other heavy metal systems.
\end{abstract}

\maketitle

\section{Introduction}
                
Precision measurements of the weak interaction, first proposed in atoms by Bouchiat and Bouchiat~\cite{BouchiatB74a,BouchiatB74b}, have been observed and are currently in progress on various parity non-conserving (PNC) transitions in numerous media including atoms, molecules, and ions. To date, the most accurate result is from the Boulder group's 1997 experiment~\cite{WoodBCMRTW97} in an atomic cesium beam measurement with a 0.35\% uncertainty. This measurement, in concert with precise theoretical models of the cesium atom~\cite{PorsevBD09,PorsevBD10,DzubaBFR12}, allows for  a precise determination of the weak charge $Q_w$.  
The theoretical efforts of Refs.~\cite{PorsevBD09,PorsevBD10,DzubaBFR12} have yielded a sub-0.5\% uncertainty calculation, and further development of an improved atomic structure model of cesium is underway~\cite{Ginges2017,Ginges2017b}.
The Boulder group's experiment also produced a measurement of the nuclear anapole moment, which results from the weak force within the nucleus~\cite{Zel1958,Flam1984}. 
Their nuclear-spin-dependent (NSD) measurement, however, 
is at odds with other measurements of the anapole moment, as discussed in Refs.~\cite{Bouchiat1991,FlambaumM97,HaxtonW01,Ginges2004}. No other significant determinations of nuclear anapole moments in atomic systems have been reported.  In short, a new measurement of PNC transitions with a lower uncertainty is needed for probing of physics beyond the standard model~\cite{MarcianoR90,HaxtonW01,Rosner02,SafronovaBDKDC18} and resolving the discrepancy the Boulder group reported in their measurement of the nuclear anapole moment. 

Several programs have recently reported exciting progress 
in high precision weak measurements. 
Antypas et al.~\cite{AntypasFSTB18} reported 0.5\% uncertainty measurements in the $6s^2~^1S_0\rightarrow5d6s^3D_1$ transition at 408 nm in four isotopes of ytterbium
to show the dependence of the weak interaction on the neutron number. Their effort to measure the weak NSD interaction continues~\cite{Antypas2017}. 
The TRIUMF collaboration~\cite{Sheng2010,AubinGBPSZCMFSOG11} have been developing techniques for trapping francium and have carried out preliminary spectroscopic measurements of this unstable alkali metal atom. Their goal is to probe the weak interaction in a chain of trapped francium isotopes. 
At Yale, the DeMille group has recently reported~\cite{AltuntasACD18a,AltuntasACD18b} progress in characterization and suppression of systematic effects in $^{138}$Ba$^{19}$F polar molecule measurements toward the weak NSD measurement in $^{137}$BaF.

Since the PNC transitions are so weak, their measurement must in each case be carried out using interference with a relatively stronger transition (e.g. magnetic dipole ($M1$), electric quadrupole ($E2$), or Stark-induced transitions). 
Optical rotation via PNC and $M1$ interference was carried out with a 1\% uncertainty in thallium ~\cite{EdwardsPBN95,VetterMMLF95} 
and in lead~\cite{MeekhofVMLF95}, and with a 2\% uncertainty in Bismuth~\cite{MacphersonZWSH91}. In atomic beam measurements, a Stark-PNC interference technique was used in cesium (e.g.~\cite{WoodBCMRTW97}) and ytterbium (e.g.~\cite{AntypasFSTB18,Antypas2017}) with modulation of net transition rates detected through fluorescence detection. 
In addition, the group of M. Bouchiat~\cite{Guena2003} has developed a pump-probe Stark-PNC interference technique for measurements in cesium where a high intensity pulse excites the forbidden transition and a moderate pulse probes the population asymmetry in the excited states via gain polarization rotation detection. This scheme has yielded 2.6\% uncertainty measurements~\cite{GuenaLB05a,GuenaLB05b}.
More recently, our group has developed a two-color coherent control scheme where an additional laser is added to ``strongly'' excite the weak transitions. This technique displayed shot-noise-limited detection in measurements of a weak $6S_{1/2}\rightarrow 8S_{1/2}$ Stark-induced transition~\cite{GunawardenaE07a,GunawardenaE07b}, and was used to measure the magnetic dipole moment $M_1$ on the $6S_{1/2}\rightarrow 7S_{1/2}$ transition~\cite{AntypasE13a,AntypasE14} in cesium. We are also working on 
a two-color optical and rf interference experiment to directly probe the NSD interaction in the cesium ground hyperfine states~\cite{Choi2016}.

The novel technique that we outline in this paper is a pump-probe gain scheme where we excite the weak transition via two-color interfering interactions with cw lasers and directly monitor the excitation rate with a cw probe field through a stimulated emission process. It involves interference between a strong two-photon and weak one-photon (Stark-induced and PNC) transitions. The primary observable in this scheme is the modulation amplitude of the probe gain signal as a function of the relative phase difference between the two-photon and one-photon transitions. 
This gain differs in several ways from that observed previously~\cite{Guena2003,GuenaLB05a,GuenaLB05b}. First, two-pathway coherent control techniques allow for direct modulation of the gain signal.  Secondly, it is not based upon the asymmetry of the population of the excited state and, hence, the observable 
is not the rotation angle of the optical polarization of the probe beam. 
And finally, our scheme involves cw rather than pulsed lasers.

The paper is organized as follows; in Sec.~\ref{sec:int}, we detail the two-color coherent control technique for a novel PNC-Stark interference measurement; in Sec.~\ref{sec:gain}, we describe the pump-probe gain scheme in cesium with reasonable experimental parameters for $6S_{1/2}\rightarrow 7S_{1/2}$ transitions; in Sec.~\ref{sec:analysis} we analyze the systematic and statistical errors; and we summarize our findings in Sec.~\ref{sec:sum}.

\begin{figure}
  \includegraphics[width=7.5cm]{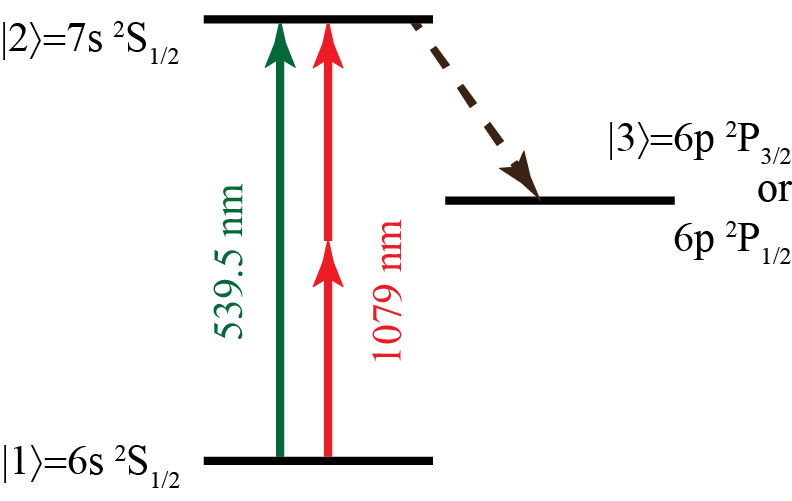}\\
  \caption{(Color online) Abbreviated energy level diagram of cesium. 
The two-color excitation lasers excite the $|1 \rangle \rightarrow |2 \rangle$ ($6s \: ^2S_{1/2} \rightarrow 7s \: ^2S_{1/2}$) transition.   
 One may interrogate the $7s$ excitation by measuring the gain in the probe laser beam, which is tuned to the $|2 \rangle \rightarrow |3 \rangle$ ($7s \: ^2S_{1/2} \rightarrow 6p \: ^2P_{3/2}$ or $7s \: ^2S_{1/2} \rightarrow 6p \: ^2P_{1/2}$) transition. 
The optical-phase-dependent population of the $7s \: ^2S_{1/2}$ state results in a modulation of the probe laser gain.   } 
  \label{fig:Cs_energy_levels_gain_exp}
\end{figure}

\section{Interfering Interactions}\label{sec:int}
We show a simplified energy level diagram of the cesium atom in Fig.~\ref{fig:Cs_energy_levels_gain_exp}.
\begin{figure*}
  \includegraphics[width=11cm]{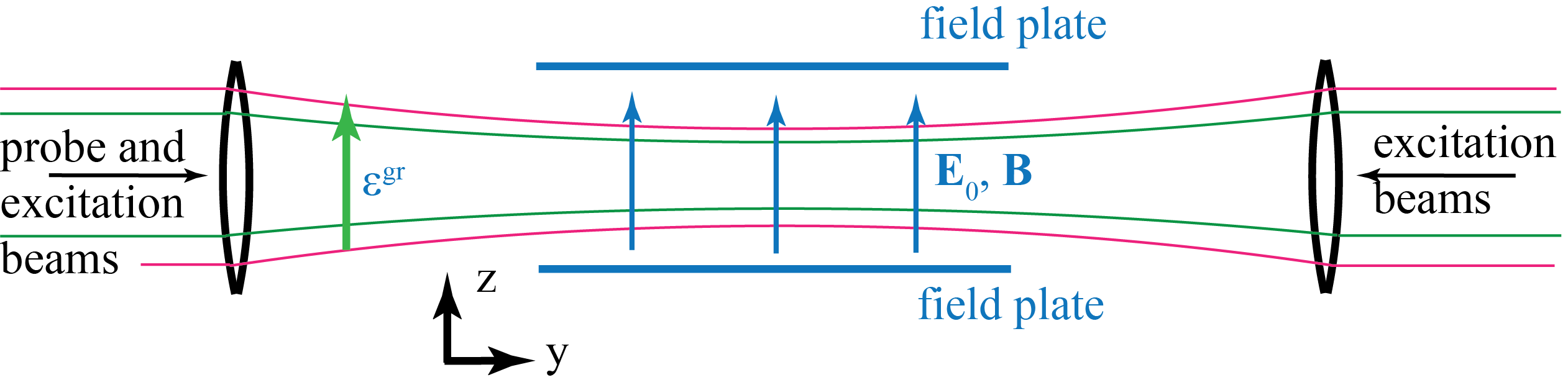}\\
  \caption{(Color online) Experimental geometry for the proposed gain measurement.  The excitation beams propagate in the $\pm y$ direction, forming a standing wave pattern, in order to reduce the magnetic dipole contributions to the excitation.  The laser polarization of the green beam, the static electric field $\mathbf{E}_0$, and a static magnetic field $\mathbf{B}$ are oriented along the $z$-direction.  The polarization of the probe beam, which propagates co-linearly with the excitation beams, can be in the $z$-direction to drive $\Delta m = 0$ transitions to the $6p \ ^2P_J$ level, or in the $x$-direction to drive $\Delta m = \pm 1$ transitions.    }
  \label{fig:exp_geom_gain}
\end{figure*}
The mutually-coherent excitation laser field components, at wavelengths of $\lambda = 1079$ nm and 539.5 nm, drive the $6s \: ^2S_{1/2} \rightarrow 7s \: ^2S_{1/2}$ transition.  We label these levels $|1 \rangle$ and $|2 \rangle$, respectively.  
The probe laser, which propagates parallel to the excitation beams, will experience gain when its frequency is resonant with the $|2 \rangle \rightarrow |3 \rangle$ transition due to the population in level $|2 \rangle$. 
State $|3 \rangle$ can be either the $6p \: ^2P_{3/2}$ level at a probe wavelength of $\lambda^{pr} = 1.47 \ \mu$m or the $6p \: ^2P_{1/2}$ level at $\lambda^{pr} = 1.36 \ \mu$m.

We have considered several different potential measurement geometries in order to evaluate their utility in this type of gain measurement.  Several requirements must be satisfied.  First, the two excitation beams (the 1079 nm beam and the 540 nm beam) must propagate co-linearly in order to maintain a constant phase difference between the various transition amplitudes for excitation of the $7s$ state throughout the interaction region.  Second, $\Delta F$ and $\Delta m$ for each of these transitions must be the same so that the amplitudes interfere with one another. ($F$ and $m$ are the quantum numbers representing the total angular momentum and its projection onto the $z$-axis, respectively).  
After consideration of the selection rules for two-photon, Stark-induced, and PNC-induced transitions, with various states of laser polarization, we have determined that the static electric field $\mathbf{E}_0$ (that is, the Stark-mixing field) must be perpendicular to the propagation direction of the excitation lasers, and that the electric field polarization $\mbox{\boldmath$\varepsilon$}^{gr}$ of the green beam (at 540 nm) must be parallel to the static field $\mathbf{E}_0$. 
We assign this direction as the $z$-direction, and show the experimental geometry in Fig.~\ref{fig:exp_geom_gain}.  For this geometry, the projection quantum number $m$ does not change for any of the excitations; that is, only $\Delta m = 0$ excitations are allowed.  Similarly, only $\Delta m = 0$ transitions are allowed in this two-photon excitation using equal frequency photons, regardless of the polarization of the 1079 nm beam~\cite{Cagnac1973}.

\begin{figure}[b!]
  \includegraphics[width=5cm]{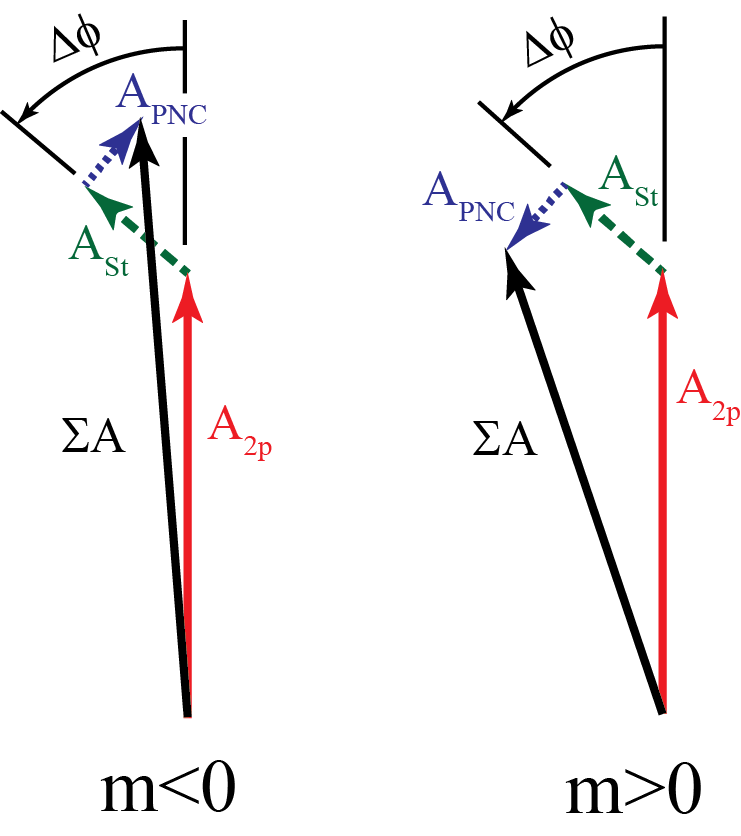}\\
  \caption{(Color online) Representations of the transition amplitudes for excitation of two hyperfine components ($m<0$ on the left, $m>0$ on the right) of the cesium $7s \: ^2S_{1/2}$ state.  The long red solid arrow is the two-photon amplitude $A_{2p}$ driven by the 1079 nm laser, while the green dashed arrow and blue dotted arrow show the Stark-induced amplitude $A_{St}$ and the PNC amplitude $A_{PNC}$, both driven by the 540 nm beam.  $\Delta \phi = 2\phi^{IR} -\phi^{gr} $ is the phase difference between the IR and green beams.  Note that the PNC amplitudes for the two diagrams are reversed, resulting in different net transition amplitudes of states $m<0$ and $m>0$.  }
  \label{fig:PhasorAmps}
\end{figure}

The total transition amplitude for excitation of the $7s \: ^2S_{1/2}$ state is the sum of amplitudes for the individual distinct interactions.  We show a representation of these amplitudes in Fig.~\ref{fig:PhasorAmps}, including the two-photon amplitude $A_{2p}$ driven by the 1079 nm laser (represented by the long, red solid arrow); a Stark-induced amplitude $A_{St}$ driven by the 540 nm beam, (the  intermediate length, green, dashed arrow); and a PNC amplitude $A_{PNC}$, also driven by the green laser (the short, blue, dotted arrow).  
For the geometry of the experiment described above, these various transition amplitudes can be written as
\begin{equation}\label{eq:A2pdef}
  A_{2p} = \tilde{\alpha} \left( \varepsilon^{IR} \right)^2 e^{-2 i \phi^{IR}}
\end{equation}
for two-photon excitation,
\begin{equation}\label{eq:AStdef}
  A_{St} = \alpha E_0 \varepsilon^{gr}  e^{- i \phi^{gr}}
\end{equation}
for Stark-induced excitation, and 
\begin{equation}\label{eq:APNCdef}
  A_{PNC} = i \Im ( \mathcal{E}_{PNC}) C_{Fm}^{Fm}  \varepsilon^{gr}  e^{- i \phi^{gr}}
\end{equation}
for the weak-force-induced amplitude, where $\mathcal{E}_{PNC}$ is the purely imaginary dipole moment induced by the weak force, and $\Im$ indicates the imaginary part.  We use the notation of Gilbert and Wieman~\cite{GilbertW86} for these transition amplitudes.  $\varepsilon^{IR}$, $\varepsilon^{gr}$, and $E_0 $ represent the field amplitudes of the 1079 nm beam, the 539.5 nm beam, and the static electric field, respectively.  We include the phases $\phi^{IR}$ and $\phi^{gr}$ of the time-varying fields, since these parameters are critical to the coherent sum of the amplitudes.  The parameter $\alpha$ is the scalar Stark polarizability (see, for example, Ref.~\cite{VasilyevSSB02}), and $\tilde{\alpha}$ is the two-photon moment.  The Stark polarizability, calculated as
\begin{eqnarray}
\alpha &=& \frac{e^2}{6} \: \sum_n \left[\rule{0in}{0.25in} \langle 7s || r || np_{1/2} \rangle  \langle  np_{1/2} || r || 6s \rangle \right. \nonumber \\
  && \left( \frac{1}{E_{7s} - E_{np_{1/2}}} + \frac{1}{E_{6s} - E_{np_{1/2}}} \right) \nonumber \\  
   && \hspace{0.4in} - \rule{0in}{0.3in} \langle 7s || r || np_{3/2} \rangle  \langle  np_{3/2} || r || 6s \rangle  \nonumber \\  && \left. \left( \frac{1}{E_{7s} - E_{np_{3/2}}} + \frac{1}{E_{6s} - E_{np_{3/2}}} \right) \right], \nonumber
\end{eqnarray}
has played a central role in the determination of the weak charge $Q_w$ of the cesium atom.  Its value (in atomic units) using the latest experimental~\cite{YoungHSPTWL94,RafacT98,RafacTLB99,AminiG03,DereviankoP02,BouloufaCD07,ZhangMWWXJ13,AntypasE13b,Tanner14,PattersonSEGBSK15} or theoretical~\cite{VasilyevSSB02,SafronovaSC16} values available for electric dipole matrix elements in cesium, is $\alpha = -264.4 \: (6) \: a_0^3$, where $a_0$ is the Bohr radius.  (To convert to SI units, divide by $6.06511 \times 10^{40}$ J$^{-1}$(V/m)$^2$.)  The two-photon moment, using the perturbation expansion for the two-photon interaction for a one-color laser beam tuned far from any one-photon interactions in Ref.~\cite{Loudon} is 
\begin{eqnarray}\label{eq:alphatilde}
  \tilde{\alpha} &=& \frac{e^2}{6} \: \sum_n \left[ \frac{\langle 7s || r || np_{1/2} \rangle  \langle  np_{1/2} || r || 6s \rangle }{ E_{np_{1/2}} - \hbar \omega} \right. \nonumber \\ 
  &&  + \left. \frac{\langle 7s || r || np_{3/2} \rangle  \langle  np_{3/2} || r || 6s \rangle }{  E_{np_{3/2}} - \hbar \omega}  \right] \nonumber
\end{eqnarray}
The numerical value of the two-photon moment in our geometry is $\tilde{\alpha} = 1006 \: (2) \: a_0^3$.  The coefficients $ C_{Fm}^{F'm'} $ come from the Clebsch-Gordon coefficients, and for the transitions of interest here are 
\begin{displaymath}
   C_{Fm}^{F'm'} = \left\{  \begin{array}{cc}
 +m/4 & \mbox{\hspace{0.5in} for $F=4$} \\ -m/4 & \mbox{\hspace{0.5in} for $F=3$}
   \end{array}  \right.
\end{displaymath}
In addition to the two-photon, Stark, and PNC moments, the $6s \: ^2S_{1/2} \rightarrow 7s \: ^2S_{1/2}$ transition is also active through a magnetic dipole interaction and an electric quadrupole interaction. The former can be suppressed (with effort and care) using counter-propagating excitation beams, as discussed in general in Ref.~\cite{BouchiatGHP82,GilbertNWW85}, and for two-pathway coherent control in particular in Ref.~\cite{AntypasE14}. The latter is not active on a $\Delta F = 0$ transition.  
We note that it will also be necessary to inhibit the two-photon Doppler-free transition, as this signal cannot interfere with the Stark-induced or weak-force induced transition.  The simplest means of doing this will be to use orthogonal polarizations for the two counter-propagating 1079 nm laser fields.   

From these expressions, one can identify the key characteristics of the phasor representations of the transition amplitudes shown in Fig.~\ref{fig:PhasorAmps}.  Under the conditions that we propose here, the two-photon amplitude $A_{2p}$ is much larger than $A_{St}$ or $A_{PNC}$.  (The relative lengths of the Stark and PNC amplitudes are magnified in Fig.~\ref{fig:PhasorAmps} for visibility.  They would be much smaller in practice.) The amplitude of $A_{St}$ is controllable, through variation of the static field strength $E_0$.  As the phase difference $\Delta \phi = 2\phi^{IR} -\phi^{gr}$ between the green and infra-red beams is varied, the phase of $A_{St} + A_{PNC}$ relative to that of $A_{2p}$ varies, and the interference can be varied between constructive and destructive.  That is, the net transition amplitude contains a large dc term (due to the two-photon amplitude alone), plus a small contribution that varies sinusoidally with phase $\Delta \phi$.  
  
When a probe laser is tuned to the $7s \: ^2S_{1/2} \rightarrow 6p \: ^2P_{J}$ transition where $J$ = 3/2 or 1/2, this probe laser will stimulate a transition to the $6p \: ^2P_{J}$ state, and will be amplified as a result.  The gain of this beam depends on the population of the $7s \: ^2S_{1/2}$ state, and varies with the magnetic component $m$, the amplitude of the electric field $E_0$, and the phase difference $\Delta \phi$.  
This gain is the basis for the measurement technique described here.

\section{Estimate of probe laser gain}
\label{sec:gain}
In this section, we will evaluate the magnitude of the gain coefficient of the probe beam resulting from the population of the $7s \ ^2S_{1/2}$ state.  When driven concurrently by the three interactions introduced above, the total excitation rate of a ground state atom to the $7s \: ^2S_{1/2}$ state is  
\begin{displaymath}
  \mathcal{R} = \frac{2 \pi}{\hbar} | A_{2p} + A_{St} + A_{PNC} |^2 \tilde{\rho}_{7s}(E),
\end{displaymath}
where $\tilde{\rho}_{7s}(E)$ is the density of states of the $7s \: ^2S_{1/2}$ state.  
On resonance, the density of states is $\tilde{\rho}_{7s}(0) =  2 / (\pi \hbar \Gamma)$, where $\Gamma$ is the decay rate of the $7s$ state, so the transition rate is 
\begin{displaymath}
  \mathcal{R} = \frac{4}{\hbar^2 \Gamma} | A_{2p} + A_{St} + A_{PNC} |^2 .
\end{displaymath}
The decay rate $\Gamma$ is $\tau_{7s}^{-1}$, where $\tau_{7s} = 48.28$ ns is the lifetime of the $7s$ state~\cite{TohJGQSCWE18}.  In steady state, the probability that an atom is in the excited state is
\begin{equation}\label{eq:rhoeqsqofsum}
  \rho_{22} = \frac{\mathcal{R}}{\Gamma} = \frac{4 }{\hbar^2 \Gamma^2} | A_{2p} + A_{St} + A_{PNC} |^2 .
\end{equation}

As shown in Meystre and Sargent~\cite{MeystreS}, the gain coefficient for the probe beam tuned to the frequency of the $|2\rangle \rightarrow |3\rangle$ transition 
\begin{equation}\label{eq:gainceffmrho32}
 \gamma = \frac{i \mu_{23} k}{\varepsilon_0} \frac{\rho_{32}}{\varepsilon^{pr}} \: n,
\end{equation} 
where $\mu_{23}$ is the electric dipole transition moment for the $|2 \rangle \rightarrow |3 \rangle$ transition, $k = 2\pi/\lambda^{pr}$ is the wavenumber of the probe beam, $\varepsilon_0 $ is the vacuum permittivity, $\rho_{32} $ is the off-diagonal matrix element for the probe transition, and $n$ is the number density of the cesium atoms.  (The sign of Eq.~(\ref{eq:gainceffmrho32}) is opposite that given in Ref.~\cite{MeystreS}, since they present the \textit{absorption} coefficient for the transition.)
In steady-state, the coherence term of the density matrix for a three-level system is 
\begin{equation}\label{eq:rho32a}
  \rho_{32} = \frac{1}	{\hbar} \frac{\mathcal{V}_{32} \mathcal{D}_{32} \left[ i \left( \rho_{33} - \rho_{22} \right) + \mathcal{D}_{31} \mathcal{V}_{12} \rho_{21} \right] }{1 + | \mathcal{V}_{12} / \hbar |^2 \mathcal{D}_{31} \mathcal{D}_{32}} ,
\end{equation}
where 
\begin{displaymath}
  \mathcal{D}_{ij} = \frac{1}{\gamma_{ij} + i \Delta_{ij}}
\end{displaymath}
describes the variation of the atomic response with detuning from resonance $\Delta_{ij}$,
\begin{displaymath}
  \mathcal{V}_{ij} = \mu_{ij} \varepsilon
\end{displaymath}
is the interaction energy with the laser field, and $\gamma_{ij}$ is the decay rate of the atomic coherence.  When collisional effects are small (as in a low-density vapor cell or in an atomic beam), one can substitute $\gamma_{ij} \rightarrow \Gamma/2$.  On resonance, therefore, where $\Delta_{ij}=0$, $\mathcal{D}_{32}$ is $\sim 2/\Gamma$.  Presuming that ($i$) the probe laser intensity is below its saturation intensity $I_{sat}^{pr}$, and ($ii$) contributions to the gain from the second (coherence) term in the numerator of Eq.~(\ref{eq:rho32a}) are negligible, the off-diagonal element $\rho_{32}$ is
\begin{equation}\label{eq:rho32b}
  \rho_{32} =  \frac{ 2 i \mu_{32} \varepsilon^{pr}}{ \hbar \Gamma} \left( \tau_{6p} \ \Sigma \Gamma_{2 \rightarrow 3}  -1 \right)  \rho_{22}.
\end{equation}
We have used the probability that an atom is in the $6p$ state following spontaneous decay from level $|2 \rangle$ is $\rho_{33} \sim \tau_{6p} (\Sigma \Gamma_{2 \rightarrow 3} ) \rho_{22}$, valid for low probe intensity.  $\tau_{6p}$ is the lifetime of the cesium $6p$ state, and $\Sigma \Gamma_{2 \rightarrow 3}$ is the total spontaneous decay rate leading to population in the hyperfine component of level $|3 \rangle$ ($6p ^2P_{1/2}$ or $6p ^2P_{3/2}$) coupled by the probe beam to level $|2 \rangle$.
The lifetime of the $6p \: ^2P_{3/2}$ state is $30.42 $ ns, and of the $6p \: ^2P_{1/2}$ state is $34.83 $ ns~\cite{YoungHSPTWL94,RafacT98,RafacTLB99,AminiG03,DereviankoP02,BouloufaCD07,ZhangMWWXJ13,PattersonSEGBSK15}. Inserting Eq.~(\ref{eq:rho32b}) for the off-diagonal matrix element into Eq.~(\ref{eq:gainceffmrho32}), the gain coefficient becomes
\begin{equation}\label{eq:gamma_rho22}
 \gamma = \frac{2 n k}{\hbar \Gamma  \varepsilon_0} | \mu_{32} |^2 \left( 1 - \tau_{6p} \ \Sigma \Gamma_{2 \rightarrow 3} \right)  \rho_{22} .
\end{equation}
\begin{figure}
  \includegraphics[width=7.5cm]{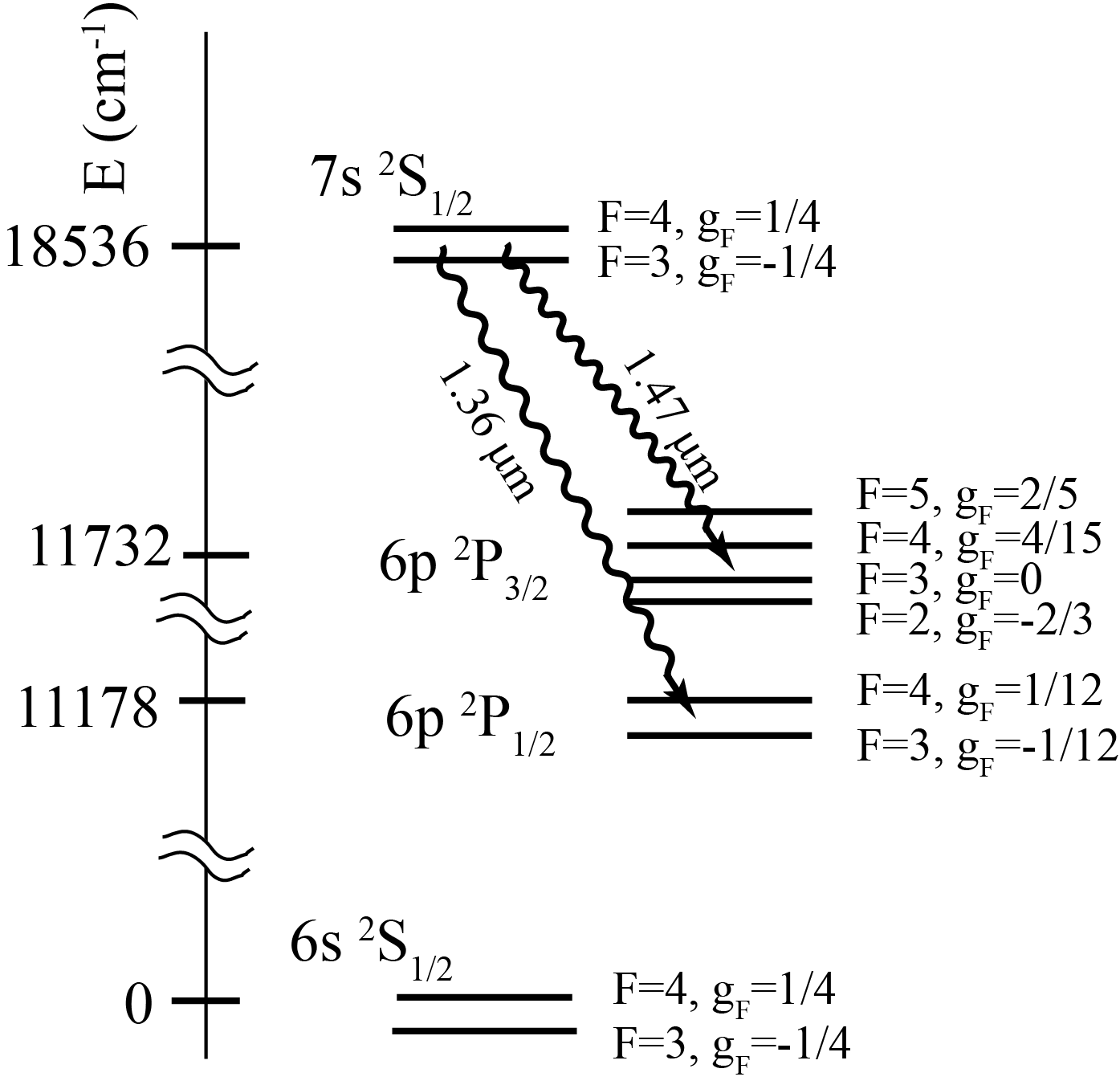}\\
  \caption{(Color online) Energy levels showing hyperfine states.}
  \label{fig:Cs_6s7s6p_levels_gain}
\end{figure}

In our multilevel system, with degenerate levels for the $6s$, $7s$, and $6p$ states, the PNC contributions to the gain coefficient from the different $m$-states tend to cancel one another.  (Due to the factor $C_{Fm}^{F^{\prime}m^{\prime}}$ in the PNC amplitude, $A_{PNC}$ for the hyperfine component $m>0$ and $A_{PNC}$ for the hyperfine component $m<0$ are opposite in sign.) 
Therefore, we need to lift the degeneracy of the individual projection states by applying a dc magnetic field $\mathbf{B} = \mathbf{a}_z B_0$ to the atoms, and measure the gain on just one Zeeman component of the probe transition. 
The Zeeman shift of each level is given by $\Delta E = - m g_F \mu_B B_0$, where $g_F$ is the Land\'{e} g-factor and $\mu_B =  9.274 \times 10^{–-24}$ J/T is the Bohr magneton. 
We label the Land\'{e} g-factors for each of the relevant levels in Fig.~\ref{fig:Cs_6s7s6p_levels_gain}.  Note that the g-factors are the same for levels $|1\rangle$ and $|2\rangle$, so the transition frequency for the $\Delta F = 0$, $\Delta m = 0$ excitation of the $7s \ ^2S_{1/2}$ state is insensitive to the application of $\mathbf{B}$. For individual probe transitions $|2\rangle \rightarrow |3\rangle$, the Zeeman shifts of the upper and lower states differ, and the transition frequencies for the individual lines are separated from one another.  

The Zeeman splitting $\Delta \nu_Z$ between adjacent lines of the probe spectrum is the least of $| g_{F^{\prime}} - g_{F^{\prime \prime}}| \mu_B B_0/h$, $2 |g_{F^{\prime \prime}}| \mu_B B_0/h  $, and $| 3g_{F^{\prime}} - g_{F^{\prime \prime}}| \mu_B B_0/h$, where $h$ is Planck's constant. We use primed notation for $7s ^2S_{1/2}$ quantities and double-primed notation for $6p ^2P_{J}$ quantities.  This splitting must be larger than the linewidth of the transition, which for the probe laser is limited by the natural linewidth of the transition, $\sim$8 MHz.  (The probe transition is Doppler free, even when the measurements are carried out in a heated vapor cell, since only atoms whose longitudinal velocity $v_y $ is zero are excited by the excitation fields when the frequency of the excitation field is tuned to line center.)  
Inserting the excited state population $\rho_{22}$ from Eq.~(\ref{eq:rhoeqsqofsum}), and assuming that the initial state population is uniformly distributed over each of the 16 hyperfine components of the ground state, the gain coefficient becomes 
\begin{equation}
 \gamma = K_0  \chi | \mu_{32} |^2  | A_{2p} + A_{St} + A_{PNC} |^2 , \label{eq:gaincoefsqrs}
\end{equation}
where the factor $K_0$ is
\begin{equation}
 K_0 =\frac{ n k }{2 (\hbar \Gamma)^3 \varepsilon_0}    \label{eq:gaincoefkappa}
\end{equation}
and 
\begin{equation}
  \chi =  1 - \tau_{6p} \ \Sigma \Gamma_{2 \rightarrow 3} .
\end{equation}

Since $A_{2p} \gg A_{St}$ and $A_{PNC}$, when we expand the square in Eq.~(\ref{eq:gaincoefsqrs}), we can drop the terms that contain $A_{St}^2$, $A_{St} A_{PNC}$, and $A_{PNC}^2$, and write 
\begin{eqnarray}
  \gamma &=& K_0  \chi | \mu_{32} |^2  
  \left\{\rule{0in}{0.17in} | A_{2p}|^2 \right. \nonumber \\ &&  \left. +  \left( A_{2p} \left[ \rule{0in}{0.13in} A_{St} + A_{PNC}  \right]^{\ast}   + c.c. \right)  \rule{0in}{0.17in}\right\}. \nonumber
\end{eqnarray}
The gain coefficient $\gamma$, which consists of three terms, depends on the polarization of the probe beam, the phase difference between the IR and green beams, and the direction and magnitude of the applied dc electric field $E_0$.  We write the gain coefficient as
\begin{displaymath}
  \gamma = \gamma_{2p} + \gamma_{St} + \gamma_{PNC},
\end{displaymath}
where $\gamma_{2p}$ is the gain coefficient due to the two-photon excitation of the 7s state alone,
\begin{displaymath}
  \gamma_{2p} = K_0  \chi | \mu_{32} |^2   | A_{2p}|^2  ,
\end{displaymath}
$\gamma_{St}$ is the gain coefficient that arises from the interference between the two-photon excitation and the Stark-induced excitation
\begin{displaymath}
  \gamma_{St} = K_0 \chi | \mu_{32} |^2  
  \left\{\rule{0in}{0.17in}  | A_{2p}| \  \alpha E_0  \varepsilon^{gr}  e^{ i \Delta \phi } + c.c.  \rule{0in}{0.17in} \right\},
\end{displaymath}
and $\gamma_{PNC}$ is the gain coefficient that arises from the interference between the two-photon excitation and the weak-force-induced excitation
\begin{eqnarray}
  \gamma_{PNC} &=& K_0 \chi | \mu_{32} |^2 \left\{\rule{0in}{0.17in} | A_{2p}|  \ i \Im ( \mathcal{E}_{PNC}) \right.  \nonumber \\ 
  & & \left. \hspace{0.7in} C_{Fm}^{F^{\prime}m^{\prime}}  \varepsilon^{gr}  e^{ i \Delta \phi } + c.c.   \rule{0in}{0.17in} \right\}. \nonumber
\end{eqnarray}

We list the transition moments, the population factor $\chi$, $ C_{Fm}^{F^{\prime}m^{\prime}}$, and the Zeeman peak separation $\Delta E_Z$ for selected lines in Table~\ref{table:moments}.
\begin{table*}
\begin{center}
    \begin{tabular}{|c|c|c|c|c|c|c|}
\hline Probe  & \rule{0.10in}{0in}Probe\rule{0.10in}{0in} & \rule{0.10in}{0in} $F^{\prime}, m^{\prime} \rightarrow F^{\prime \prime}, m^{\prime \prime}$ \rule{0.10in}{0in} &  $\displaystyle \left|  \frac{\mu_{32}}{e \langle 7s || r || 6p_{J} \rangle} \right |^2$  &  \rule{0.15in}{0in} \rule{0in}{0.2in} $ \chi $ \rule{0.15in}{0in} &  \rule{0.10in}{0in} \rule{0in}{0.2in} $ C_{Fm}^{F^{\prime}m^{\prime}}$ \rule{0.10in}{0in} &  \rule{0.10in}{0in} $\Delta \nu_Z / B_0  \rule{0.10in}{0in} $ \\ transition & pol. & & & & & (MHz/G)  \\
\hline \hline
	& &  \rule{0in}{0.2in}$4, 4 \rightarrow 5, 5$	&  1/4 = 0.250 & $0.796$ & +1 & 0.210 \\  
$7s \: ^2S_{1/2} \hspace{0.15in}$	& horiz. &  $4, 3 \rightarrow 4, 4$	& \rule{0.15in}{0in} 7/240 = 0.029\rule{0.15in}{0in}  & $0.881$ & +3/4 & 0.023  \\  
\hspace{0.15in} $\rightarrow 6p \: ^2P_{3/2}$  &	&  $3, 3 \rightarrow 2,2$	&  5/28 = 0.179 & $0.796$ & -3/4 & 0.117 \\  
\cline{2-7}
(1.47 $\mu$m)	& vert.	&  \rule{0in}{0.2in}$4, 4 \rightarrow 4, 4$	&  7/60 = 0.117 & $0.881$ &  +1  & 0.023  \\  
&  & $3,3 \rightarrow 3,3$	&  9/64 = 0.141 & $0.847$ & -3/4 & 0.350   \\  
	 
\hline \hline
$7s \: ^2S_{1/2}\hspace{0.15in} $	& horiz. &   $4, 4 \rightarrow 3, 3$&  7/24 = 0.292  & $0.810$ &  \rule{0in}{0.2in} +1 & 0.467   \\  
\hspace{0.15in} $\rightarrow 6p \: ^2P_{1/2}$	& &  $3,3 \rightarrow 4,4$	&  7/24 = 0.292 & $0.852$ &  -3/4 & 0.233   \\  
\cline{2-7}
(1.36 $\mu$m)	& vert.	&  \rule{0in}{0.2in}$4, 4 \rightarrow 4,4 $	&  1/6 = 0.167 & $0.894$ & +1 &  0.233   \\  	 
\hline 
    \end{tabular}
\caption{Squares of transition moments for several possible probe transitions.    
We also list the population factor $\chi$, the angular momentum coefficients $ C_{Fm}^{F^{\prime}m^{\prime}}$ and the Zeeman peak separation between magnetic components $\Delta \nu_Z / B_0  $ for each of these lines.}\label{table:moments}
\end{center}
\end{table*} 

As an example, we consider in detail the gain on the $7s \: ^2S_{1/2}, \: (4, \ 4)   \rightarrow 6p \: ^2P_{3/2}, \: (5, \ 5)$ probe transition.  (The numbers enclosed within the parentheses are $F$ and $m$ for the two states.)  The PNC gain is largest on this line, and the Zeeman peak separation, while not maximal, is sufficient.  The individual gain coefficients are
\begin{equation}\label{eq:gamma2p}
  \gamma_{2p} = K_0  \chi \left( \frac{ e^2}{4}  | \langle 7s || r || 6p_{3/2} \rangle |^2 \right)   | A_{2p}|^2  ,
\end{equation}
\begin{eqnarray}\label{eq:gammaSt}
  \gamma_{St} &=& K_0 \chi \left( \frac{ e^2}{4}  | \langle 7s || r || 6p_{3/2} \rangle |^2  \right) \nonumber \\
   && \times | A_{2p}| \alpha E_0   \varepsilon^{gr}  2 \cos( \Delta \phi )   ,
\end{eqnarray}
and 
\begin{eqnarray}
  \gamma_{PNC} &=& K_0 \chi  \left( \frac{ e^2}{4}  | \langle 7s || r || 6p_{3/2} \rangle |^2 \right) \nonumber \\ 
& &  \hspace{-0.1in} \times | A_{2p}|   \Im ( \mathcal{E}_{PNC})   \varepsilon^{gr} (-2) \sin( \Delta \phi ) .\label{eq:gammaPNC}
\end{eqnarray}
Each of these gain coefficients contains the factor 
\begin{displaymath}
   K_0 e^2 |\langle 7s || r || np_{j} \rangle |^2 =  \frac{ n k e^2 }{2 (\hbar \Gamma)^3 \varepsilon_0}  |\langle 7s || r || np_{j} \rangle |^2 , \nonumber 
\end{displaymath} 
so we will start by evaluating this.  We use 
\begin{itemize}
  \item $n = 3.7 \times 10^{12}$ cm$^{-3}$.  We derive this effective density using the equilibrium vapor density of cesium at a temperature of 180$^{\circ}$C ($8.3 \times 10^{14}$ cm$^{-3}$), reduced by the factor $\Delta \nu_n / \Delta \nu_D$ for the $6s \rightarrow 7s$ transition, where $\Delta \nu_n$ is the natural linewidth ($\Delta \nu_n$ = 3.3 MHz) for the transition and $\Delta \nu_D$ is its Doppler width ($\Delta \nu_D$ = 750 MHz).    
    \item $k = 2\pi/\lambda^{pr} = 4.27 \times 10^4 $ cm$^{-1}$ for the 1.47 $\mu$m probe beam and $4.62 \times 10^4 $ cm$^{-1}$ for the 1.36 $\mu$m probe beam.
  \item $\Gamma = \tau_{7s}^{-1} = 2.1 \times 10^7$ s$^{-1}$~\cite{TohJGQSCWE18}.
    \item $\langle 7s || r || 6p_{3/2} \rangle = 6.487 \: a_0$ and $\langle 7s || r || 6p_{1/2} \rangle = 4.245 \: a_0$, as determined from the lifetime  of the $7s$ state $\tau_{7s} = 48.28$ ns~\cite{TohJGQSCWE18} and presuming the ratio of transition moments is $ \langle 7s || r || 6p_{3/2} \rangle / \langle 7s || r || 6p_{1/2} \rangle = 1.528$~\cite{DzubaFKS89,BlundellSJ92,DzubaFS97,SafronovaJD99}. 
\end{itemize}
The common factor $K_0 e^2 |\langle 7s || r || np_{j} \rangle |^2$ is 
\begin{equation}\label{eq:multfacjeq3halves}
   K_0 e^2 |\langle 7s || r || 6p_{3/2} \rangle |^2 = 2.6 \times 10^{59} \: {\rm J}^{-2}{\rm m}^{-1}
\end{equation}
for the 1.47 $\mu$m line, and 
\begin{equation}\label{eq:multfacjeq1half}
   K_0 e^2 |\langle 7s || r || 6p_{1/2} \rangle |^2 = 1.2 \times 10^{59} \: {\rm J}^{-2}{\rm m}^{-1}
\end{equation}
for the 1.36 $\mu$m line.  These pre-factors show that the gain is typically larger on the $7s \: ^2S_{1/2} \rightarrow 6p \: ^2P_{3/2}$ probe transition than on the $7s \: ^2S_{1/2} \rightarrow 6p \: ^2P_{1/2}$ line.

Finally, we must determine the transition amplitudes $A_{2p}$, $A_{St}$, and $A_{PNC}$.  Using $\tilde{\alpha} = 1006 \: a_0^3$, we determine the two-photon transition amplitude $A_{2p} = \tilde{\alpha} \left( \varepsilon^{IR} \right)^2 $ and the transition rate $\mathcal{R}_{2p}$.  Assuming the power of the 1079 nm beam as 5 W, and a Gaussian beamshape with the radius of the beam as $w^{IR} \sim $ 1.0 mm, the field amplitude on the axis is
\begin{displaymath}
  \varepsilon^{IR} = 4.9 \times 10^4 \: {\rm V/m}. 
\end{displaymath}
Then the two photon amplitude, from Eq.~(\ref{eq:A2pdef}), is 
\begin{equation}\label{eq:A2pval}
  A_{2p}/\hbar = 3.8 \times 10^5 \: {\rm s}^{-1},
\end{equation}
the two-photon excitation rate $\mathcal{R}_{2p}$ alone (that is, without the Stark or PNC contributions) is
\begin{displaymath}
  \mathcal{R}_{2p} = \frac{4}{\Gamma} \left(\frac{ A_{2p}}{\hbar} \right)^2 = 2.8 \times 10^4 \: {\rm s}^{-1}, 
\end{displaymath}
and the net probability of finding an atom in the excited state is
\begin{displaymath}
  \mathcal{R}_{2p} \tau_{7s} \simeq 0.0013.
\end{displaymath}
As required to avoid saturation effects, this probability is much less than unity.

We use the values for $K_0 e^2 |\langle 7s || r || np_{j} \rangle |^2 $ (Eq.~(\ref{eq:multfacjeq3halves})) and $A_{2p}$ (Eq.~(\ref{eq:A2pval})) to find the on-axis gain coefficient $\gamma_{2p}$.  
\begin{equation}\label{eq:gamma2pval}
  \gamma_{2p} = 81 \: {\rm m}^{-1} . \nonumber
\end{equation}
We define the gain factor due to the two-photon excitation alone as 
\begin{displaymath}
  G_{2p} \equiv \gamma_{2p} \ell_{\rm gain}   = 8.1,
\end{displaymath}
where we use a vapor cell length $\ell_{\rm gain}$ = 10 cm, assuming good beam overlap between the probe beam and the excitation beams over the full length.

We next evaluate the gain coefficient $\gamma_{PNC}$ which comes from the interference between the two-photon interaction and the PNC interaction.  For this, we need the PNC moment $\Im ( \mathcal{E}_{PNC})$ and the field amplitude $\varepsilon^{gr}$ of the green beam.  (The product of these two terms gives us the amplitude $|A_{PNC}|$ of the $6s \: ^2S_{1/2}, \:  (4, \ 4)  \rightarrow 7s \: ^2S_{1/2}, \: (5, \ 5)$  component.)  The calculated value of $\mathcal{E}_{PNC}$ is about $i 0.9 \times 10^{-11} (-Q_w/N) \: ea_0$, where $(Q_w \sim - 76$ is the weak charge and $N = 78$ is the neutron number of cesium~\cite{PorsevBD09,PorsevBD10,DzubaBFR12}.  (This is converted to SI units using $e a_0 = 8.4735 \times 10^{-30}$ Cm.)  Presuming a green laser power of 3 W and a beam radius of $w^{gr} = w^{IR}/\sqrt{2}$, the field amplitude of the green beam is
\begin{displaymath}
  \varepsilon^{gr}  = 5.4 \times 10^4 \: {\rm V/m}. 
\end{displaymath}
The PNC amplitude for the $m=4$ level is then 
\begin{equation}\label{eq:APNC}
 |A_{PNC}/\hbar| = \Im ( \mathcal{E}_{PNC})  \varepsilon^{gr} /\hbar \approx 0.038  \: {\rm s}^{-1}. 
\end{equation}
Using Eqs.~(\ref{eq:gammaPNC}, \ref{eq:multfacjeq3halves}, \ref{eq:A2pval}, and \ref{eq:APNC}), the amplitude of the gain coefficient $\gamma_{PNC} $ is therefore
\begin{equation}
  |\gamma_{\rm PNC}| = 1.6 \times 10^{-5} \: {\rm m}^{-1} , \nonumber 
\end{equation}
and the gain factor is 
\begin{displaymath}
  G_{\rm PNC} = |\gamma_{\rm PNC}| \ell_{\rm gain}  = 1.6 \times 10^{-6}.
\end{displaymath}

The gain $G_{\rm St}$ due to the Stark-induced amplitude is variable, since this gain depends on the applied dc electric field $E_0$.  In a measurement, one would apply a field $E_0$ that produces a gain $G_{\rm St}$ comparable to the PNC gain.  Therefore we rely on our evaluation of $G_{\rm PNC}$, and identify the range of values for $G_{\rm St}$ between zero and $\sim 4 \times G_{\rm PNC}$.

\section{Measurement Scheme}
\label{sec:analysis}
\begin{figure}
  \includegraphics[width=8.5cm]{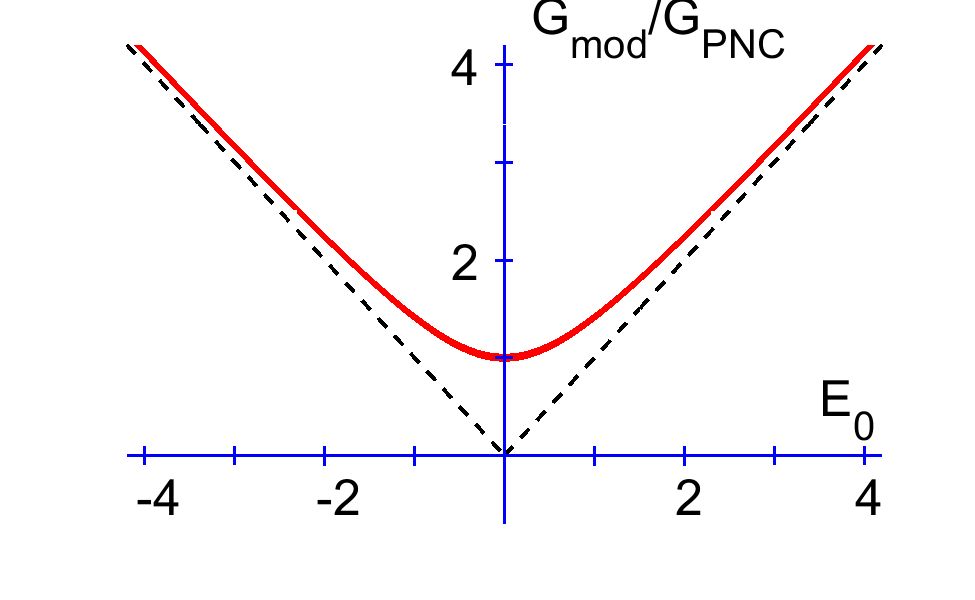}\\
  \caption{(Color online) A plot of the amplitude of the modulating gain $G_{\rm mod}$, normalized by $G_{\rm PNC}$, versus the static electric field strength $E_0$.  The field amplitude is in units of $\mathcal{E}_{PNC}/\alpha$.}
  \label{fig:Mod_Amp_vs_E}
\end{figure}

In a measurement of the gain in this system, we need to be able to separate the three gain contributions $G_{\rm 2p}$, $G_{\rm St}$, and $G_{\rm PNC}$ from each other, and to use the Stark gain to calibrate the measurement of the PNC gain.  In our proposed scheme, we examine the gain for a horizontally-polarized probe beam passing through the gain region, tuned to the selected Zeeman-shifted hyperfine component of the gain transition.  In this geometry, one could detect the increase in the probe power, and exploit the following signatures of the three contributions to the gain to differentiate them.  The two-photon gain $G_{2p}$ does not depend on $\Delta \phi$, the phase between the green and IR beams, while $G_{St}$ and $G_{PNC}$ do.  And while $G_{St}$ and $G_{PNC}$ each modulate sinusoidally with $\Delta \phi$, these gains are $\pi/2$ out of phase with each other,
\begin{eqnarray}
  I_{\rm out}^{pr} &=& I_{\rm in}^{pr} \left( e^{G_{\rm 2p} + G_{\rm St}\cos(\Delta \phi) - G_{\rm PNC}\sin(\Delta \phi)} \right)^2 \nonumber \\ &  \approx & I_{\rm in}^{pr}  e^{2 G_{\rm 2p}} \left( \rule{0in}{0.15in} 1 + 2 G_{\rm St}\cos(\Delta \phi) \right. \nonumber \\ 
  & & \hspace{0.6in}  \left. - 2G_{\rm PNC}\sin(\Delta \phi) \rule{0in}{0.15in} \right).\nonumber
\end{eqnarray}
Therefore, the amplitude of the sinusoidally-varying modulation of the net gain is the quadrature sum of these individual gain terms, \begin{equation}
  G_{\rm mod} = \sqrt{G_{\rm St}^2 + G_{\rm PNC}^2}.
\end{equation}  
That is, the gain modulation amplitude grows hyperbolically with the static field strength $E_0$, as we show in Fig.~\ref{fig:Mod_Amp_vs_E}, and measurements of $G_{\rm mod}$ vs.\ $E_0$ can yield $\mathcal{E}_{PNC}/\alpha$.  This is a feature that we have designed into our previous coherent control schemes as well. 
A complete determination requires repeated measurements at each field strength $E_0$, and measurements at several different values of $E_0$; in all, perhaps 20--25 measurements.

We can estimate the signal-to-noise ratio (S/N) of such a measurement as follows.  We assume that the intensity of the probe beam after amplification by the gain medium is one-tenth the saturation intensity $I_{\rm sat}$ of the probe transition, which we estimate as 
\begin{displaymath}
  I_{\rm sat} \approx \frac{c \varepsilon_0\Gamma^2}{2  e^2 |\langle 7s | r | 6p_{j} \rangle |^2} \sim 0.2 \: {\rm mW/cm}^2
  \end{displaymath}
for the $7s \: ^2S_{1/2}, \: (4, \ 4) \rightarrow 6p \: ^2P_{3/2}, \: (5, \ 5)$ probe line, where $c$ is the speed of light in vacuum.  (The probe beam intensity can influence the precision of the measurement, since  saturation effects can suppress the gain.  High precision measurements will require a probe intensity much less than $I_{\rm sat}$, but due to the gain within the vapor cell, saturation effects are not an issue at the entrance of the cell, but only towards the exit.  The limiting input intensity is difficult to estimate in advance.)  This beam will overlap and be parallel to the excitation beams; we choose a probe beam radius that matches that of the 540 nm beam, $w^{\rm pr} = w^{\rm gr}$, so the power of the probe beam after amplification in the vapor cell is $P_{out}^{pr} = I_{in}^{pr} e^{2G_{\rm 2p}} \pi (w^{pr})^2/2 = 0.16 \: \mu {\rm W}$.  The photodiode signal $S$ that one would measure is 
\begin{displaymath}
  S = S_{PD} P_{out}^{pr} \left( 2 \times G_{PNC} \right),
\end{displaymath}
where $S_{PD}$ is the sensitivity of the photodiode used to measure the probe beam power (0.9 A/W for InGaAs photodiode from Hamamatsu, for example).  This yields 
\begin{displaymath}
  S =  0.5 \times 10^{-12} \: {\rm A}.
\end{displaymath}

The noise of the measurement will be primarily shot noise arising from the large dc component of the probe beam power $P_{out}^{pr}$.  This property is a characteristic of heterodyne detection, for which the signal consists of a large dc term (in this case due to two-photon absorption) alone, plus a modulation term due to the interference between the strong (two-photon amplitude) and the weak (Stark and/or PNC amplitudes) term.  We estimate the level of the shot noise using
\begin{displaymath}
  i_N^2 = 2 e I \Delta \nu,
\end{displaymath}
where $e = 1.6 \times 10^{-19} \: {\rm C}$ is the fundamental charge unit, $I = S_{PD} P_{out}^{pr} $ is the average photo-current generated by the photodetector, and $\Delta \nu$ is the bandwidth of the detector.  The signal-to-noise ratio is
\begin{displaymath}
    S/N = \frac{ 2  S_{PD} P_{out}^{pr} G_{PNC} }{\sqrt{2 e (S_{PD} P_{out}^{pr} ) \Delta \nu}} = \sqrt{ \frac{ 2 S_{PD} P_{out}^{pr}}{ e  \Delta \nu}} \: G_{\rm PNC}.  
\end{displaymath}     
Using the parameters we have discussed above, the S/N ratio is 
\begin{equation} \label{eq:SNratio}
  S/N \sim 2.3\sqrt{t(s)},
\end{equation}
where $t = 1/\Delta \nu$ is the integration time of a measurement.  
For example, to achieve a S/N ratio of 100 in a single measurement, an integration time of 21 minutes is required. 

While direct comparison of the S/N ratio between measurements is difficult because of the many different parameters used, the potential of the proposed gain measurement technique is promising. The Bouchiat group's pump-probe gain polarization rotation experiment~\cite{Guena2003,GuenaLB05a,GuenaLB05b} yielded an S/N of $\sim0.9\sqrt{t(s)}$ with a reasonable pulse repetition rate and optical density. (We arrived at this expression using the equation in Section F of Ref.~\cite{GuenaLB05a}.) The Boulder group's atomic beam measurement in 1997~\cite{WoodBCMRTW97,Wood1999} yielded an S/N ratio of $\sim 0.6\sqrt{t(s)}$, and the ytterbium measurements by Antypas et al.~\cite{AntypasFSTB18} reached a $S/N \sim 0.55 \sqrt{t(s)}$. Both of these measurements employ a power build-up cavity to enhance the field amplitude.  We emphasize that we derived Eq.~(\ref{eq:SNratio}) for the present gain measurement for the case of no power build-up cavity. Further improvement in the S/N is possible with a dual-wavelength (1079 nm and 539.5 nm) power build-up cavity in our gain measurement scheme. A fine-temperature-controlled vapor cell with a minimal reflection loss at the windows~\cite{Jahier2000} can be placed inside a high finesse cavity to increase the PNC gain by several times, which may enhance the S/N by the same factor.
Such a cavity geometry may complicate the experimental setup, but we concluded that no additional systematic effects would be introduced due to the cavity.

It is interesting to speculate as to whether this probe gain detection technique can be competitively applied to an atomic beam experiment.  
We have evaluated the S/N ratio for our current experiment, replacing our current detection scheme based upon fluorescence detection on a cycling transition such as described in Refs.~\cite{WoodBCMRTW97,AntypasE13a} with a detection scheme based on probe gain, but keeping all other aspects of the experiment, such as beam density, the size and power of the excitation laser, the same, and find that the S/N ratio of a PNC measurement using the probe gain technique can be as much as twice that of the fluorescence detection method.

Atomic parity violation experiments are notorious for their sensitivity to systematic errors introduced by stray uncontrolled dc electric and magnetic fields and small alignment imperfections between $\mathbf{B}$, $\mathbf{E}_0$, and  $\mbox{\boldmath$\varepsilon$}^{gr}$.  We have previously reported on the expected systematics 
for this geometry of the excitation beams in Ref.~\cite{AntypasE14}.  In that work, we showed that many systematic effects encountered in previous measurements are reduced or eliminated in the two-pathway coherent control schemes through the use of ($i$) low amplitude static electric fields, and ($ii$) modulation of the frequency of the phase difference $\Delta \phi$, combined with phase-sensitive detection. 
As shown in Sec. IV of Ref.~\cite{AntypasE14}, the primary systematics that are expected in the geometry of the proposed measurement are due to stray static electric field components in the $x$-direction $\Delta E_x$ of the form $\alpha \Delta E_x (\varepsilon_x^{gr})^{\prime \prime}$, and $M_1 (\varepsilon_x^{gr})^{\prime \prime}$, where $(\varepsilon_x^{gr})^{\prime \prime}$ is the imaginary component of the 540 nm beam in the $x$-direction. (This component exists only when the optical polarization is slightly elliptical.)  Reduction of the former requires weak stray fields $\Delta E_x$, while the latter is reduced by using counter-propagating beams of equal intensity.  Both effects are reduced by using a highly-linearly-polarized 540 nm laser beam.  We estimate that these unwanted contributions can reasonably be controlled to levels less than $10^{-4}$ of the PNC contribution.

\section{Summary}
\label{sec:sum}
In this report, we have proposed and analyzed a new scheme in which one may detect the gain in a probe laser beam in two pathway coherent control to determine the amplitude of the parity non-conserving weak-force-induced electric dipole amplitude.  While we have considered specifically the cesium $6s \ ^2S_{1/2} \rightarrow  7s \ ^2S_{1/2}$ transition in this work, with the probe beam tuned to the $7s \ ^2S_{1/2} \rightarrow  6p \ ^2P_{3/2}$ transition, the probe laser could in principle be tuned to any $7s \ ^2S_{1/2} \rightarrow  np \ ^2P_{3/2}$, where $n > 6$, as well.  Our choice of $n=6$ is guided by three factors: ($i$) For $n=6$, the probe beam is amplified, rather than attenuated as it would be for $n > 6$. The larger probe beam power after amplification aids in the detection sensitivity.  ($ii$) The dipole moment for the transition to $n=6$ is larger, leading to relatively large gain.  ($iii$) The wavelength for the probe transition to the $6p \ ^2P_{3/2}$ can be generated with a commercially-available diode laser.  

We also suggest that the probe gain technique discussed in this paper could in principle be applied to PNC investigations in other heavy atomic systems. For instance, the TRIUMF collaboration has taken steps to measure the PNC amplitudes in rubidium in parallel with a measurement in a francium MOT~\cite{Sheng2010}. Our technique can be easily adopted for the PNC measurement in a rubidium vapor cell, since it is the second heaviest stable alkali metal.

\section{Acknowledgement}

This material is based upon work supported by the National Science Foundation under Grant Numbers PHY-1607603 and 1404419-PHY.
 We would like to acknowledge fruitful discussions with Marie-Anne Bouchiat, F. Robicheaux, Jun Ye, T. S. Zwier, J. S. M. Ginges, M. Safronova, and Jay Meikrantz at Precision Glassblowing.

\bibliography{biblio}

\end{document}